# A single-layer panoramic metalens with > 170° diffraction-limited field of view

**Mikhail Y. Shalaginov[1,*], Sensong An[2], Fan Yang[1], Peter Su[1], Dominika Lyzwa[3], Anuradha Agarwal[1,4], Hualiang Zhang[2], Juejun Hu[1,*], and Tian Gu[1,4*]**

[1]Department of Materials Science & Engineering, Massachusetts Institute of Technology, Cambridge, MA 02139, USA

[2]Department of Electrical & Computer Engineering, University of Massachusetts Lowell, Lowell, MA 01854, USA

[3]Department of Biological Engineering, Massachusetts Institute of Technology, Cambridge, MA 02139, USA

[4]Materials Research Laboratory, Massachusetts Institute of Technology, Cambridge, MA 02139, USA

*[mys@mit.edu](mys@mit.edu), [hujuejun@mit.edu](hujuejun@mit.edu), [gutian@mit.edu](gutian@mit.edu)

**Abstract**

Wide-angle optical functionality is crucial for implementation of advanced imaging and image projection devices. Conventionally, wide-angle operation is attained with complicated assembly of multiple optical elements. Recent advances in nanophotonics have led to metasurface lenses or metalenses, a new class of ultra-thin planar lenses utilizing subwavelength nanoantennas to gain full control of the phase, amplitude, and/or polarization of light. Here we present a novel metalens design capable of performing diffraction-limited focusing and imaging over an unprecedented > 170° angular field of view (FOV). The lens is monolithically integrated on a one-piece flat substrate and involves only a single layer of metasurface that corrects third-order Seidel aberrations including coma, astigmatism, and field curvature. The metalens further features a planar focal plane, which enables considerably simplified system architectures for applications in imaging and projection. We fabricated the metalens using Huygens meta-atoms operating at 5.2 μm wavelength and experimentally demonstrated aberration-free focusing and imaging over the entire FOV. The design concept is generic and can be readily adapted to different meta-atom geometries and wavelength ranges to meet diverse application demands.

**Wide-angle metalenses: state-of-the-art**

Wide-angle optical systems are vital to high performance imaging, detection, image or beam projection, and Fourier optics, among many other applications[1–4]. One of the earliest examples of such systems is the panoramic camera pioneered by Thomas Sutton in the year 1858, which consisted of a single water-filled spherical lens producing an image on a curved glass plate covered with reactive emulsion. Due to apparent difficulties in fabrication and handling of curved plates, the original approach was soon abandoned but it outlines the fundamental challenges associated with achieving wide-FOV imaging. Since then panoramic photography has been evolving along the path of planar detector planes while relying on compound lens assemblies, commonly known as "fisheye lenses", to reduce optical aberrations at large field angles. Such multi-lens architecture, however, increases the size, weight, and assembly complexity of optical systems.

Metasurfaces, devices capable of controlling the phase and amplitude of propagating light with arrays of subwavelength structures, present a promising solution enabling flat and compact optical components[5–14]. Metasurface-based designs have been widely employed for constructing planar ultra-thin lenses, also called metalenses[15–22], to mitigate several types of aberrations[23], in particular spherical[16] and chromatic[24–27] aberrations. However, angle-dependent aberrations (e.g. coma, astigmatism, and field curvature) are among the major challenges that must be overcome to realize optical systems with enhanced functionalities while maintaining a minimum element count[28,29]. The prevailing method for designing a single-element metalens uses a hyperbolic phase profile to realize a spherical wavefront[16]:

$$\phi_{ideal} = -\frac{2\pi}{\lambda}\left(\sqrt{f^2 + x^2 + y^2} - f\right) \quad (1)$$

where $\lambda$ is the wavelength of incident light, $x$ and $y$ are the coordinates of meta-atoms, and $f$ is the focal length of the metalens. While generating zero spherical aberration at the focal plane for a planar wavefront at normal incidence, such a phase profile is not optimized for obliquely incident beams. When a beam strikes the metasurface at an oblique incident angle $(\theta_x, \theta_y)$, the desired phase profile becomes:

$$\phi_{oblique} = \frac{2\pi}{\lambda}\left\{\sqrt{f^2 + [x - f\tan(\theta_x)]^2 + [y - f\tan(\theta_y)]^2} - [x\sin(\theta_x) + y\sin(\theta_y)]\right\} \quad (2)$$

The deviation between the two phase distributions due to different angles of incidence (AOIs) results in third-order (Seidel) aberrations such as coma, astigmatism, and field curvature, which limit the FOV. As an example (see Supplementary Information), assuming a baseline metalens design with 1 mm diameter and 2 mm focal length operating at 5.2 μm wavelength, the conventional hyperbolic phase profile effectively suppresses spherical aberration and achieves diffraction-limited focusing with a unity Strehl ratio at normal incidence. However, at AOIs larger than 7°, coma aberration becomes dominant, reducing the Strehl ratio to below 0.8 and rapidly degrading the metalens performance from the diffraction limit. The small viewing angle significantly limits the use of a single metalens in imaging and image projection applications.

Several metalens designs have already been implemented to suppress coma and expand the diffraction-limited FOV. One approach entails engraving metasurfaces on spherical surfaces, which however poses a non-trivial fabrication challenge[23]. Another solution involves cascading multiple metasurfaces based on traditional bulk optical system design principles. In such doublet metalens designs[28,29], the focusing function is primarily performed by one of the metasurfaces while the other acts to correct the off-axis wavefront aberrations. Diffraction-limited FOV up to approximately 56° was demonstrated in such doublets[28]. In comparison, the FOV of single-layer metalens has been limited to 30° with reduced diffraction limit resolution due to vignetting[30], and the design further suffers from low optical efficiencies of 6-20% and sensitivity to assembly misalignment (Table 1). Metalenses with wide-angle performance rivaling their traditional refractive counterparts have not been realized to date.

Table 1. Optical metalenses with diffraction-limited wide-angle performance

| | Diffraction-limited FOV (°) | Focusing efficiency (%) | Number of metasurface layers | Wavelength (nm) | Numerical aperture |
|---|---|---|---|---|---|
| Arbabi, et al.[28] | 56 | 45-70 | 2 | 850 | 0.49 |
| Groever, et al.[29] | 50 | 30-50 | 2 | 470-660 | 0.44 |
| Engelberg, et al.[30] | 30 | 6-20 | 1 | 825 | 0.2 |
| This work | > 170 | 32-45 | 1 | 5200 | 0.24 |

In this paper, we demonstrate the first panoramic metalens with a record diffraction-limited FOV exceeding 170°. The wide field-of-view (WFOV) metalens assumes a simple and easy-to-fabricate configuration, consisting of only a single metasurface layer and an aperture co-integrated on a single thin substrate. Moreover, the lens has a planar focal plane which significantly simplifies the associated detector (for imaging and detection) or light emitter (for image/beam projection, display, etc.) array design. Here we experimentally implemented the design in the mid-infrared (mid-IR) using Huygens meta-atoms[31–39], although the design is completely generic and scalable to other meta-atom structures and wave bands. The metalens design concept, fabrication approach and characterization results are described in the following sections.

**Extreme wide FOV metalens: the concept**

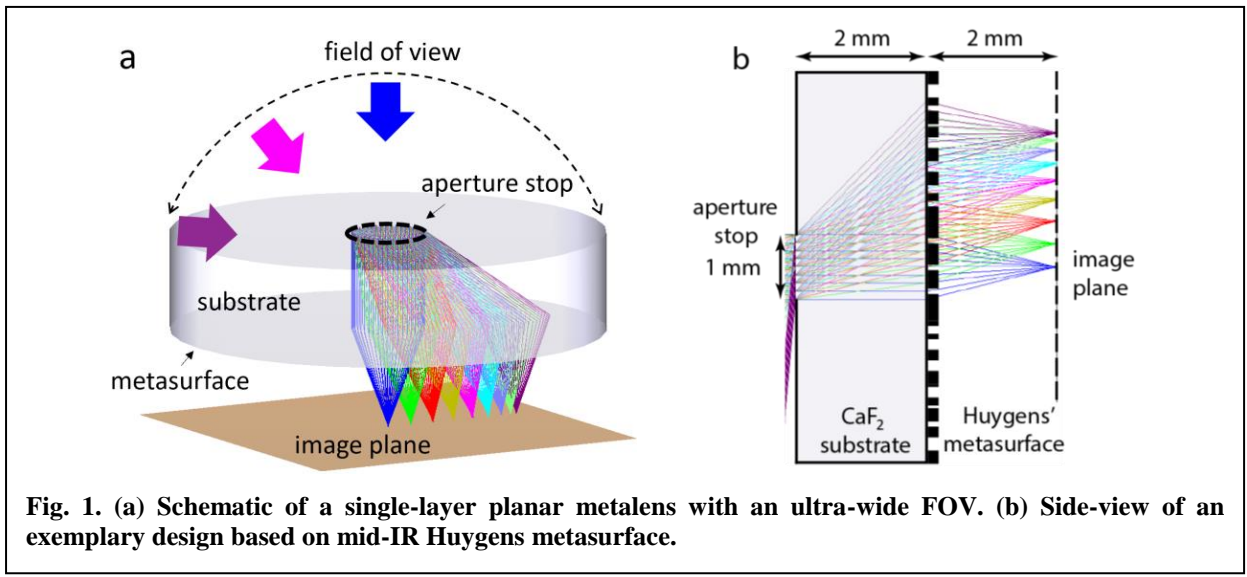

**Fig. 1. (a) Schematic of a single-layer planar metalens with an ultra-wide FOV. (b) Side-view of an exemplary design based on mid-IR Huygens metasurface.**

The basic concept of the WFOV metalens is schematically illustrated in Fig. 1a. It consists of a single substrate with an input aperture positioned on the front surface and a metasurface positioned on the back surface. The substrate has a refractive index of $n_{sub}$ and a thickness of $t_{sub}$. Light beams incident on the input aperture at different incident angles $\theta_i$ are refracted to the backside metasurface and then focused onto a planar focal plane.

By spatially decoupling the metasurface and aperture stop while positioning them on the same substrate, such a metalens architecture allows input beams at different AOIs to be captured on different yet continuous portions of the metasurface, facilitating local optimization of the phase profiles. The metasuface phase profile is designed so that the RMS wavefront error from an ideal spherical wavefront over the input aperture is always smaller than 0.07 wavelength. This ensures that a Strehl ratio over 0.8 can be maintained over the entire field-of-view[23], thereby achieving diffraction-limited performance at various light illumination conditions.

It is important to distinguish our approach from spatially multiplexed designs[40,41], where non-overlapping regions on a metasurface are dedicated to beams at different AOIs. As a result, such metalenses can only achieve high quality focusing for a discrete set of incident angles. In our case, the judiciously designed metasurface phase profile and metalens architecture allow diffraction-limited focusing of beams with continuously varying incident angles despite mutually overlapping beam profiles on the metasurface. Therefore, our metalens can achieve aberration-free beam focusing or conversely, beam collimation and hence image projection for any light direction from or to any point on the front hemisphere.

In addition to correcting aberrations such as coma and astigmatism, the metalens features a planar focal plane across the entire FOV. The elimination of Petzval field curvature presents a critical benefit to imaging and image projection applications by facilitating standard planar detector or emitter array integration.

Schematic of an exemplary WFOV metalens design operating at 5.2 µm wavelength is shown in Fig. 1b. A 2-mm-thick calcium fluoride (CaF$_2$) planar substrate ($n_{sub}$ = 1.4 at 5.2 µm) is used. A 1-mm-diameter circular aperture is positioned on the substrate front side and a 5.2 × 5.2 mm$^2$ metasurface is patterned on the back side. The metasurface contains 2,000 × 2,000 Huygens meta-atoms made of PbTe with a square lattice constant of 2.5 µm. The metasurface is designed to have a constant focal length of 2 mm, corresponding to an effective numerical aperture (NA) of 0.24. At an incident angle of nearly 90°, the maximum angle of light propagation inside the substrate is 45.7°. As shown in the next section, the phase response of our meta-atoms is only weakly dependent on the beam incident angles within this range.

**Device design and modeling**

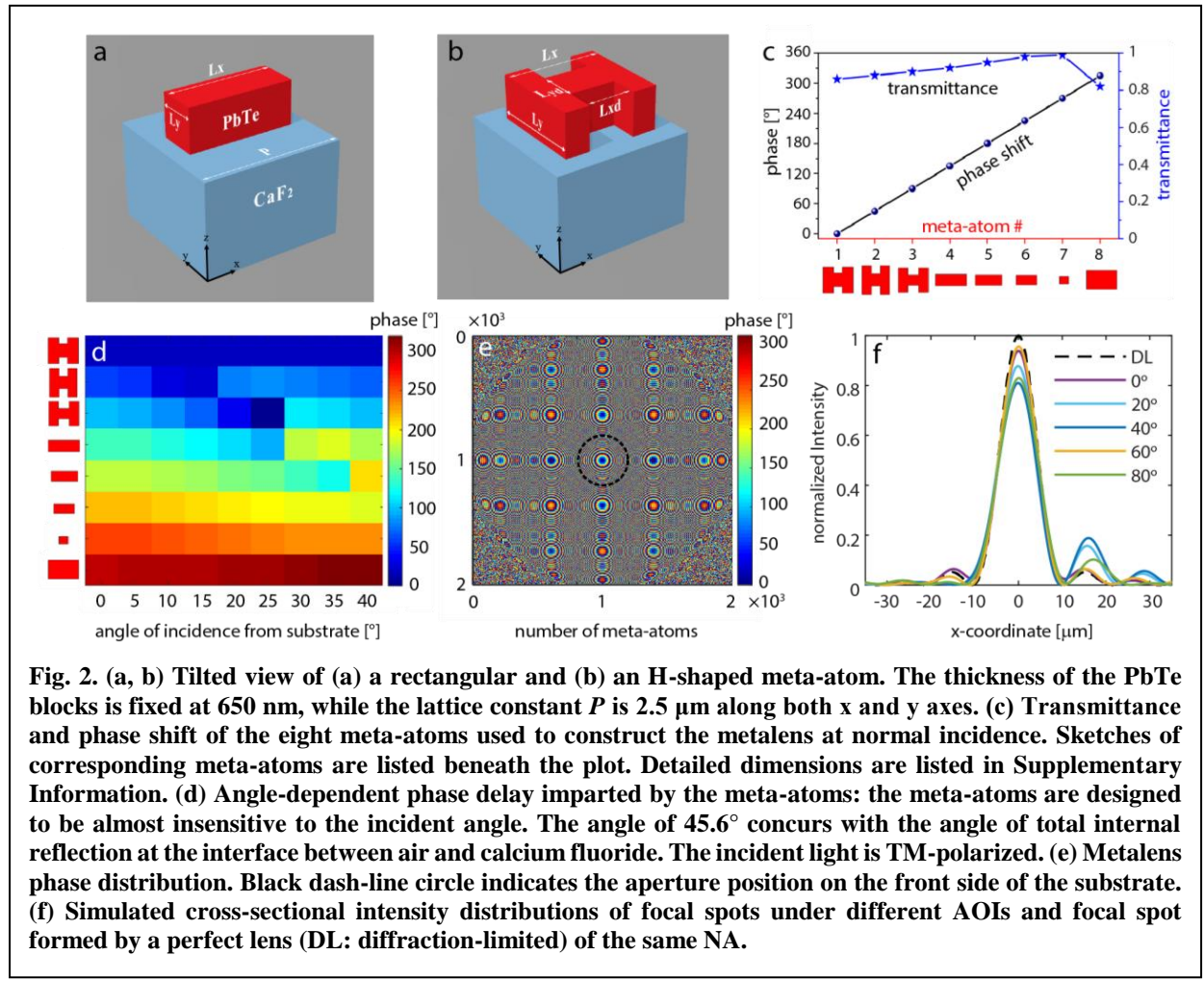

Fig. 2. (a, b) Tilted view of (a) a rectangular and (b) an H-shaped meta-atom. The thickness of the PbTe blocks is fixed at 650 nm, while the lattice constant $P$ is 2.5 µm along both x and y axes. (c) Transmittance and phase shift of the eight meta-atoms used to construct the metalens at normal incidence. Sketches of corresponding meta-atoms are listed beneath the plot. Detailed dimensions are listed in Supplementary Information. (d) Angle-dependent phase delay imparted by the meta-atoms: the meta-atoms are designed to be almost insensitive to the incident angle. The angle of 45.6° concurs with the angle of total internal reflection at the interface between air and calcium fluoride. The incident light is TM-polarized. (e) Metalens phase distribution. Black dash-line circle indicates the aperture position on the front side of the substrate. (f) Simulated cross-sectional intensity distributions of focal spots under different AOIs and focal spot formed by a perfect lens (DL: diffraction-limited) of the same NA.

The metalens was designed utilizing a hierarchical combination of finite element method (FEM) and Kirchhoff diffraction integral as described in Methods. At the sub-wavelength-scale, full wave FEM simulations were implemented to design and model the meta-atoms for desired optical responses. At the macroscopic system level, the diffraction integral method incorporating the full wave simulation results enables computationally efficient validation of the focusing characteristics of the entire metalens and was used to optimize the phase profile of the metasurface (see Supplementary Information for further details).

The Huygens meta-atoms comprise rectangular or H-shaped blocks made of PbTe resting on a $CaF_2$ substrate as illustrated in Figs. 2a and 2b[34]. The material combination is chosen to take advantage of their low optical losses and giant refractive index contrast in the mid-IR spectral range, enabling metasurface operation in the transmissive mode while supporting both electric and magnetic dipole (ED & MD) resonances. Their shapes were designed to obtain spectrally overlapping ED and MD resonances at the operation wavelength, conducive to full 360° (2π) phase coverage with near-unity transmittance leveraging the Kerker effect[35]. The meta-atom library consists of eight elements covering the 360° phase space with a discrete step of 45° for linearly TM-polarized light at 5.2 μm wavelength. The amplitude and phase responses of each meta-atom simulated at normal incidence are illustrated in Fig. 2c. The meta-atom dimensions are listed in Supplementary Information Table S2. The phase shift of each meta-atom at oblique AOIs (inside the substrate) was also simulated and summarized in Fig. 2d. The results indicate that the meta-atom responses are only weakly dependent on incident angle (see Supplementary Information for further details).

An analytical model based on the Kirchhoff diffraction integral is utilized to analyze the full metasurface performance under different AOIs. The model incorporates angular-dependent phase masks following individual meta-atom responses under different AOIs obtained from full wave simulations (Supplementary Information, Fig. S4). The optimized metalens phase profile is shown in Fig. 2e and further detailed in Supplementary Information. Compared to the ideal phase profile, the RMS wavefront errors of the designed phase profile at all AOI values (Eq. 2) are consistently less than 0.07 wavelength, leading to Strehl ratios better than 0.8. As a result, when compared to a perfect lens with the same NA and focal length, the metalens design achieves diffraction limited focusing and imaging performance across the entire FOV (Fig. 2f). Modulation transfer functions (MTF) simulated at different AOIs are shown in Fig. S5 to further support this conclusion.

## Metalens fabrication and characterization

The metalens was fabricated using electron beam lithography on a 2-mm-thick $CaF_2$ planar substrate by a double-resist-layer lift-off method following previously published protocols (see Methods for full details)[34]. The meta-atoms were made of thermally evaporated nanocrystalline PbTe and have a uniform thickness of 650 nm. The frontside aperture was subsequently defined by a metallic tin layer using standard UV lithography. Figure 3 shows an SEM top-view micrograph of the fabricated metasurface, confirming good pattern fidelity consistent with our design.

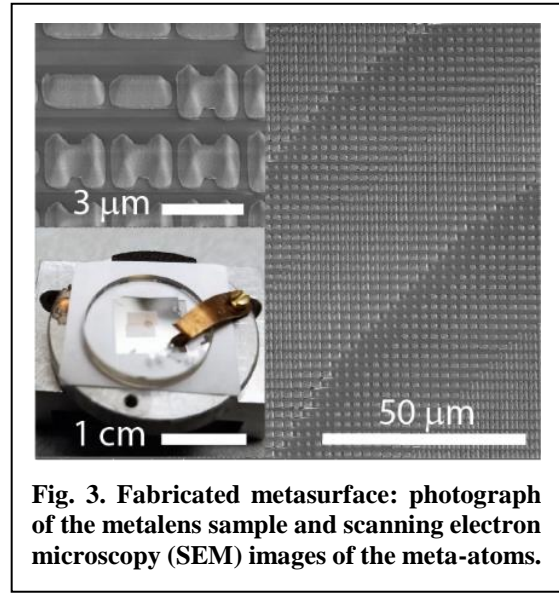

**Fig. 3. Fabricated metasurface: photograph of the metalens sample and scanning electron microscopy (SEM) images of the meta-atoms.**

We started with characterizing the focal spot quality of the lens at various AOIs. In the experiment, the sample was illuminated by a collimated, linearly polarized laser beam from the frontside (the aperture side) at 5.2 μm wavelength. The laser was mounted on a circular rail allowing variation of the beam incident angle from 0° to 85° (Fig. 4a). The maximum incident angle of 85° was limited by geometric constraints of our experimental setup rather than the lens performance. The focal spot image was magnified using a pair of mid-IR lenses with a calibrated magnification of 120 ± 3 and projected onto a liquid nitrogen cooled InSb focal plane array (FPA) camera. Several examples of the focal spot images are presented in Figs. 4b-g, and the cross-sectional optical intensity profiles of the focal spots at 0°, 70°, and 85° incident angles are plotted in Fig. 4h inset alongside the simulated ideal focal spot profiles from an aberration-free lens with the same NA for comparison. We further computed the Strehl ratios from the measurement following previously established procedures (Fig. 4h)[34,42]. For all the incident angles, the Strehl ratio remains above 0.8, indicating diffraction-limited focusing performance of the metalens.

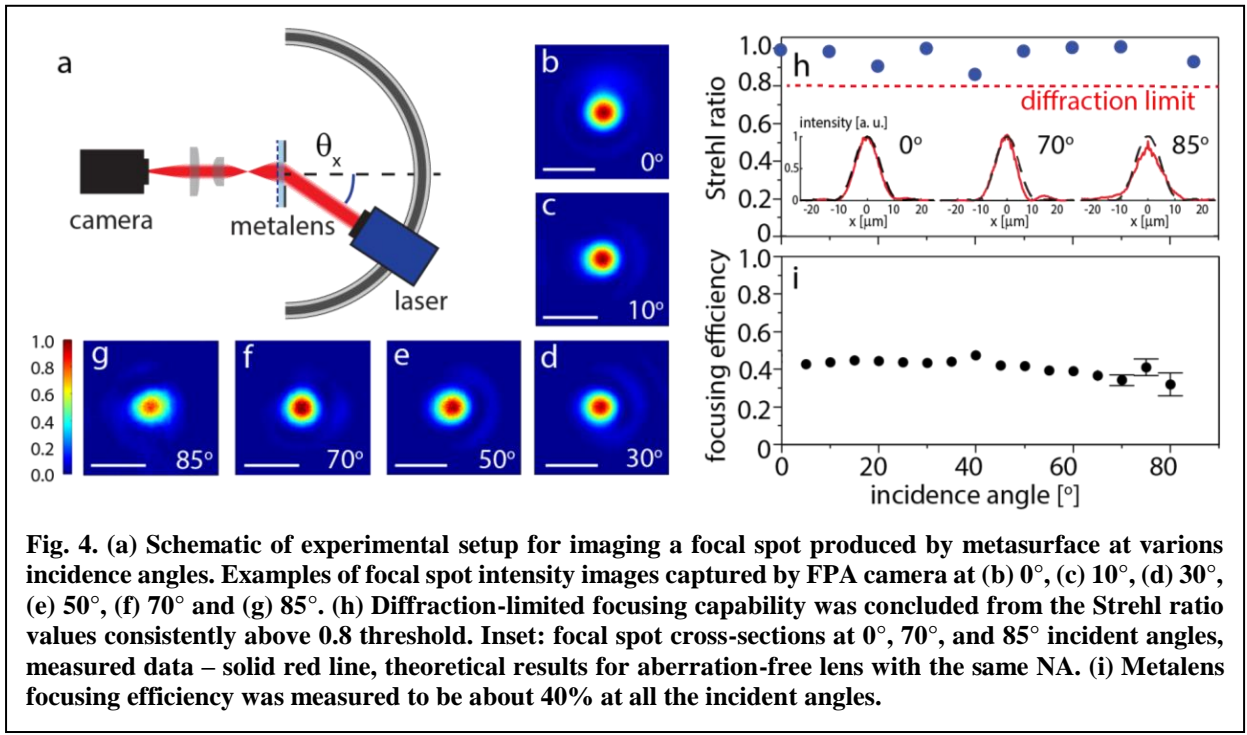

**Fig. 4. (a)** Schematic of experimental setup for imaging a focal spot produced by metasurface at various incidence angles. Examples of focal spot intensity images captured by FPA camera at **(b)** 0°, **(c)** 10°, **(d)** 30°, **(e)** 50°, **(f)** 70° and **(g)** 85°. **(h)** Diffraction-limited focusing capability was concluded from the Strehl ratio values consistently above 0.8 threshold. Inset: focal spot cross-sections at 0°, 70°, and 85° incident angles, measured data – solid red line, theoretical results for aberration-free lens with the same NA. **(i)** Metalens focusing efficiency was measured to be about 40% at all the incident angles.

We then quantified focusing efficiency of the lens, which is defined as the ratio between the power confined at the focal spot and the power incident onto the metasurface. Details of the measurement protocols are furnished in Methods. Figure 4i presents the measured focusing efficiency at different AOIs for linearly polarized light. The result indicates a relatively weak dependence on the incident angle varying from 45% to 32% as the angle of incidence changes from 0° to 85°. This relatively flat angular response is a useful feature in providing nearly uniform illumination across an image formed by the metalens.

**Metalens imaging demonstration**

To demonstrate the wide-angle imaging capability, we used a setup depicted in Fig. 5a, where the metalens collects the light scattered by an object and projects it onto the InSb FPA camera through a mid-IR lens. In the experiment, the distance between the object and the lens is fixed to 2 mm to be consistent with the planar geometry of the lens focal plane. The object consists of metallic tin patterns replicating the USAF resolution test chart. The selected test target pattern (Group 5, Element 1) contains three stripes, each 15.6 μm wide, close to the ideal diffraction-limited resolution of the lens (13.2 μm). Figure 5b shows clearly resolved images of the pattern recorded at the full angular range of our experimental setup from 0° to 82° (the measurement range is similarly bound by geometric constraints of our experimental setup). The result confirms diffraction-limited imaging performance of the metalens over a record wide angular regime.

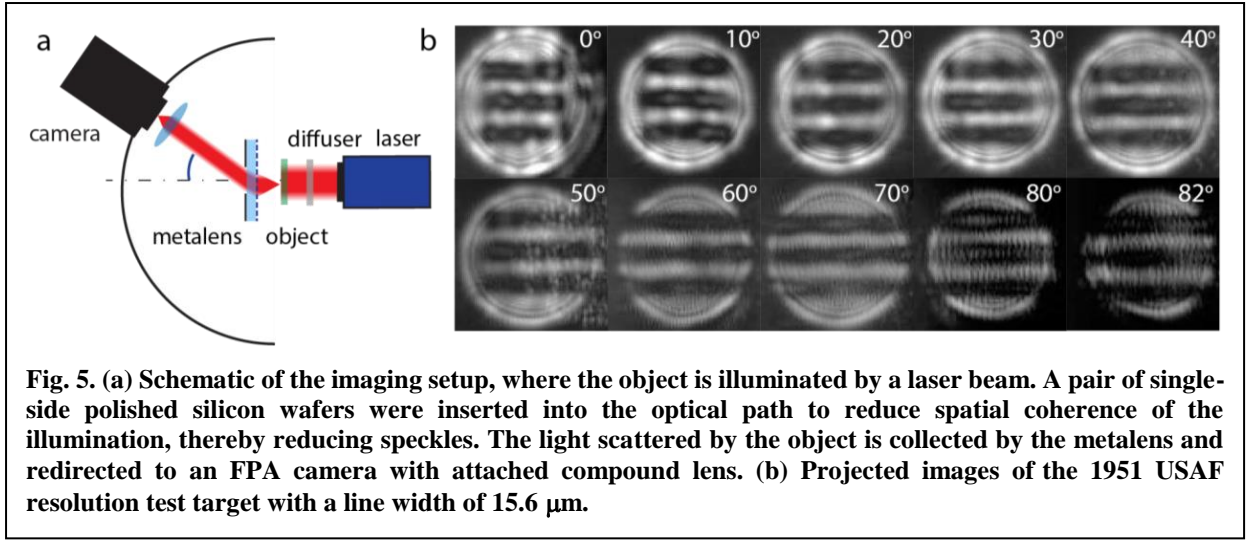

**Fig. 5. (a) Schematic of the imaging setup, where the object is illuminated by a laser beam. A pair of single-side polished silicon wafers were inserted into the optical path to reduce spatial coherence of the illumination, thereby reducing speckles. The light scattered by the object is collected by the metalens and redirected to an FPA camera with attached compound lens. (b) Projected images of the 1951 USAF resolution test target with a line width of 15.6 μm.**

**Discussion and conclusion**

In our experiment, we have chosen to embody the WFOV metalens design using Huygens metasurfaces operating in the mid-IR. The implementation leverages our previously demonstrated PbTe-on-CaF$_2$ meta-atom platform with established benefits including exceptionally high index contrast conducive to high-quality-factor Mie resonances, an ultra-thin meta-atom profile, and ease of fabrication. The choice of a Huygens metasurface structure at the same time imposes constraints such as sensitivity to polarization and wavelength (our prior work has shown diffraction-limited performance across ~ 300 nm wavelengths in the mid-IR, corresponding to 6% fractional bandwidth[34]). We want to highlight that the wide-FOV design principle described herein is generic and applicable to arbitrary meta-atom configurations. With proper choice of meta-atoms, our design can also lead to panoramic metalenses with on-demand characteristics including broadband operation and polarization diversity.

To showcase versatility of our approach, we present a polarization-insensitive near-infrared (NIR) metalens design with diffraction-limited performance covering almost an entire hemispherical view. Figure 6a illustrates an exemplary WFOV metalens design operating at 940 nm wavelength and with an effective NA of 0.2. The metasurface consists of amorphous Si (a-Si) nano-posts patterned on the back surface of a sapphire (Al$_2$O$_3$) substrate and an aperture stop positioned on the front side of the substrate. The nano-post diameters are changed to create varying phase delay covering 0 to $2\pi$[43–46]. The metalens and metasurface designs are detailed in Supplementary Information. Maximum AOI at the input aperture (90° at the air/Al$_2$O$_3$ interface) corresponds to an AOI of 34.2° inside the substrate at the metasurface interface. Full-wave simulations using the finite-difference time-domain (FDTD) method (Lumerical FDTD Solutions) suggest that phase responses of the nano-post meta-atoms are independent of the beam incident angles over the entire AOI range and the optical responses are also polarization insensitive (Figs. S7 & S8).

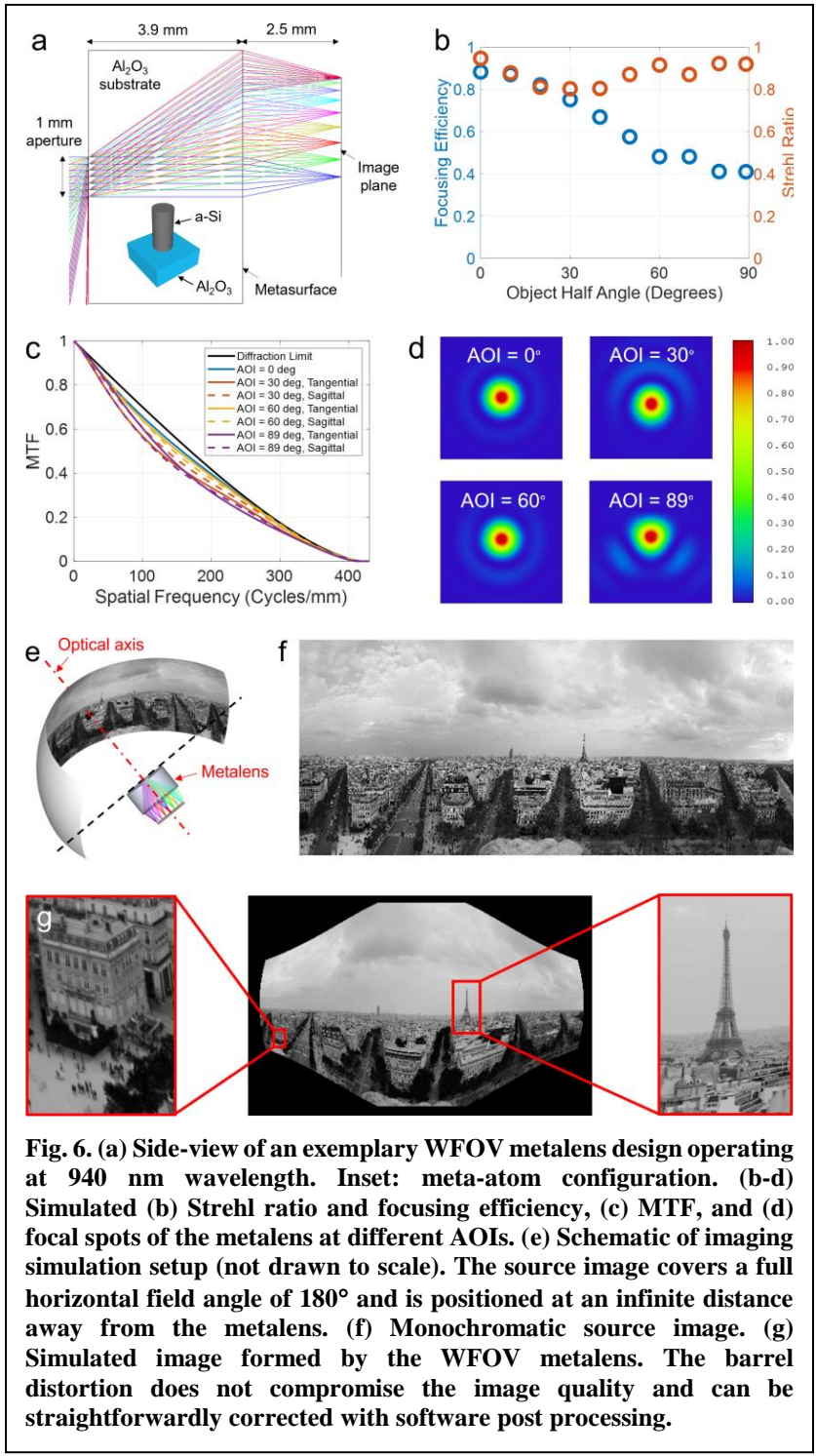

**Fig. 6. (a)** Side-view of an exemplary WFOV metalens design operating at 940 nm wavelength. Inset: meta-atom configuration. **(b-d)** Simulated (b) Strehl ratio and focusing efficiency, (c) MTF, and (d) focal spots of the metalens at different AOIs. **(e)** Schematic of imaging simulation setup (not drawn to scale). The source image covers a full horizontal field angle of 180° and is positioned at an infinite distance away from the metalens. **(f)** Monochromatic source image. **(g)** Simulated image formed by the WFOV metalens. The barrel distortion does not compromise the image quality and can be straightforwardly corrected with software post processing.

The focusing performance of the metalens under different AOIs is evaluated using Kirchhoff diffraction integral with the meta-atom phase and amplitude responses obtained from full-wave simulations. As shown in Fig. 6b, the simulated Strehl ratios are over 0.8 across the entire FOV, indicative of aberration-free focusing performance. The focusing efficiency of the metasurface versus incident angles (Fig. 6b) closely follows the average transmittance response of the meta-atoms (Fig. S7b), indicating minimum phase error induced at both the meta-atom and metasurface levels.

Imaging performance of the NIR WFOV metalens is numerically investigated using OpticStudio® (Zemax, LLC). Figures 6c and 6d plot the MTF and focal spot intensities of the metalens simulated at different AOIs. Our WFOV metasurface design achieves a diffraction-limited spatial resolution of 2.9 μm on the image plane and an average angular resolution of approximately 0.1° across the entire 180° FOV. The Image Simulation tool in OpticStudio® is further utilized to numerically evaluate the metalens' imaging performance taking into full account aberration and diffraction effects. We use a monochromatized, stitched 180° horizontal panoramic photo of Paris taken from the Arc de Triomphe de l'Étoile as the scene[47]. Figure 6e illustrates the imaging configuration and details of the model are available in Supplementary Information. Figure 6g shows the simulated image captured by the WFOV metalens. Despite the apparent barrel distortion common to wide-angle optical systems, the WFOV metalens can readily image the scene from all angles with high quality and minimal aberrations. The distortion can be straightforwardly corrected with post processing and the angular resolution can be further improved by increasing the aperture size while keeping the same NA. This design example reinforces that our WFOV design concept is generic and can be readily adapted to different meta-atom geometries and wavelength ranges to meet diverse application demands.

In conclusion, we proposed and demonstrated a novel metalens architecture to enable ultra-wide-angle panoramic imaging and image projection. Our approach uniquely combines an unprecedented hemispherical FOV, diffraction-limited performance over the entire viewing field, a remarkably compact and simple configuration involving only one metasurface layer on a flat substrate, and a planar focal plane ideal for optical system integration. As a proof-of-concept, we validated the design in the mid-IR wave band using Huygens metasurfaces, experimentally realizing a metalens with a record diffraction-limited FOV exceeding 170°. The same meta-optic architecture can be readily adapted to other wavelength ranges and meta-atom platforms. These critical advantages foresee potential integration of the technology in next-generation systems for imaging, optical projection, augmented reality/virtual reality, beam steering, and 3-D depth sensing applications.

## Methods

**Metasurface fabrication.** The sample was fabricated on a circular $CaF_2$ substrate (Edmund Optics) with a diameter of 15 mm and a thickness of 2 mm. Given inherent symmetry of the metasurface layout, only a 2 mm × 3.6 mm section of the metasurface was needed and fabricated to validate the WFOV performance. Prior to fabrication, the substrate surface was cleaned in sequential acetone and isopropanol alcohol (IPA) sonication baths for 3 minutes each. Afterwards the sample was baked at 190 °C for 5 minutes to fully evaporate solvent and adsorbed moisture on its surface. Then the substrate was treated with oxygen plasma (150 W, 1 minute, pressure 0.8 Torr) to remove organic residue contaminants. One side of the sample was covered with a double-layer photoresist composed of PMGI (800 nm thick) and ZEP 520A (400 nm thick). The PMGI layer was spin-coated at 2400 revolutions per minute (rpm) for 1 minute, then baked at 190°C for 3 minutes. The baking step is critical for assuring mechanical stability of the PMGI layer. The ZEP layer was spin-coated at 4000 rpm for 1 minute and baked at 190°C for 2 minutes. To prevent charging effects while performing electron beam (e-beam) lithography, we covered the sample with a water-soluble conductive polymer (ESpacer 300Z, Showa Denko America, Inc.) and placed a conducting clamp on top of the substrate[48]. The metasurface patterns were written with an e-beam lithography system (Elionix ELS F-125) at a voltage of 125 kV, current 10 μA, and proximity effect correction (PEC) with a base dose of 380 μC/cm$^2$. The ZEP layer was developed by submerging the sample into water, ZEDN50, and IPA for 1 minute each. The PMGI layer was subsequently partially dissolved with RD6 developer diluted in a 1:1 ratio with water. This step must be done carefully in order to achieve a necessary undercut and not to collapse the pattern. After photoresist development, a 650-nm-thick PbTe film was deposited by thermal evaporation in a custom-designed system (PVD Products, Inc.) at a rate of 17 Å/s and a base pressure of $10^{-6}$ Torr[49,50]. Before deposition, the sample was pre-cleaned with oxygen plasma to improve adhesion of the film. Later the metasurface pattern was transferred by lifting off the material on top of the photoresist by overnight soaking in N-Methyl-2-pyrrolidone (NMP). On the other side of the sample we patterned a circular aperture of 1 mm in diameter. The side patterned with the PbTe metasurface was protected by a dry film photoresist (DuPont MX5000 series) during the aperture fabrication. To fabricate the aperture, the surface was cleaned with oxygen plasma, spin-coated with a negative photoresist NR9-1000PY (Futurrex, Inc.) at 1500 rpm for 1 minute. The sample was soft baked at 115 °C, exposed to UV light through the mask for 40 s (Karl Suss MA6 Mask Aligner), and hard-baked at 155 °C. The exposed photoresist was subsequently developed in the aqueous developer RD6 (Futurrex, Inc.) for 10 s and rinsed with deionized water afterwards. Then a 200-nm layer of metal tin was deposited by thermal evaporation and lifted off by removing the photoresist with acetone. Finally, the dry film photoresist covering the metasurface side was removed by overnight NMP treatment.

**Metalens characterization.** For focal spot characterization, the metalens sample was mounted on a 3-axis translation stage and illuminated from the aperture side with a 5.2 μm collimated laser beam (Daylight Solutions, 21052-MHF-030-D00149). The laser was mounted on a mobile cart that can move along a custom-made circular rail. Focal spot produced by a metalens was magnified with a custom-made microscope assembly (henceforth termed as magnifier), consisting of lens 1 (C037TME-E, Thorlabs Inc.) and lens 2 (LA8281-E, Thorlabs Inc.). Finally, magnified image of the focal spot was projected onto a liquid nitrogen cooled InSb FPA with 320 × 256 pixels (Santa Barbara Infrared, Inc.). Magnification of the microscope assembly was calibrated to be 120 ± 3 with a USAF resolution chart. The FPA and magnifier were fixed on a breadboard, therefore both of them can be controllably translated as a single piece perpendicular to the metalens optical axis. Incident beam angle was varied from 0° to 85° with an increment of 5°. Geometric constraints due to the circular rail and the mobile cart limits the maximum measurement angle to 85°.

To demonstrate imaging capability of the metalens, we used stripe patterns from the USAF resolution test chart. The chosen pattern (group 5, element 1) consists of three stripes, each 15.6 μm wide spaced by 15.6 μm. In the imaging setup the camera had to be rotated, since the compound lens (Asio lens 40494-AA1, f 25mm, Janos Tech) FOV is limited to 45°. Wide-angle imaging performance can alternatively be achieved by introducing a second large-FOV metalens to refocus the light incident at large angles into the camera image sensor focal plane. In the latter case, a rotating camera is not necessary.

Details of the metalens focusing efficiency measurement are described in Supplementary Information.

**Device modeling.** The meta-atom simulations were carried out with a frequency domain solver in the commercial software package CST Microwave Studio. For each meta-atom, unit cell boundary conditions were employed at both negative and positive $x$ and $y$ directions, while open boundary conditions were set along the $z$-axis. Each meta-atom was illuminated from the substrate side with an $x$-polarized plane wave pointing towards the positive $z$ direction. The results shown in Fig. 2c are the phase and amplitude of the complex transmission coefficient derived between the two open ports placed at the top and bottom of each meta-atom.

The focusing and imaging behavior of the WFOV meta-lens was modeled following the Kirchhoff diffraction integral, a physically rigorous form of the Huygens-Fresnel principle[51]. The model starts with computing the Huygens point spread function of the optical system. It incorporates angular-dependent phase profiles at the metasurface and propagates wavefronts emitted from each meta-atom with corresponding amplitude and phase to the image plane where its complex amplitude is derived. The diffraction of the wavefront through space is given by the interference or coherent sum of the wavefronts from the Huygens sources. The intensity at each point on the image plane is the square of the resulting complex amplitude sum.

## Acknowledgments

This work was funded by Defense Advanced Research Projects Agency Defense Sciences Office (DSO) Program: EXTREME Optics and Imaging (EXTREME) under Agreement No. HR00111720029. The authors also acknowledge characterization facility support provided by the Materials Research Laboratory at Massachusetts Institute of Technology (MIT), as well as fabrication facility support by the Microsystems Technology Laboratories at MIT and Harvard University Center for Nanoscale Systems. The views, opinions and/or findings expressed are those of the authors and should not be interpreted as representing the official views or policies of the Department of Defense or the U.S. Government.

## Author contributions

T.G. conceived the metalens concept. M.Y.S. fabricated the devices. T.G. and S.A. designed and modeled the devices. M.Y.S. and T.G. performed device characterizations. T.G., M.Y.S., and S.A. analyzed the experimental data. P.S. contributed to device fabrication. F.Y. and D.L. contributed to optical testing. M.Y.S., T.G., and J.H. drafted the manuscript. J.H., T.G., H.Z. and A.A. supervised and coordinated the research. All authors contributed to revising the manuscript and technical discussions.

## Competing financial interests

The authors declare no competing financial interests.